\documentclass[a4paper]{article}
\usepackage{ae}
\usepackage[english]{babel}
\usepackage{latexsym}
\usepackage{amssymb,amsmath}
\usepackage{epsf,graphicx}
\newtheorem{prop}{Proposition}[section]
\newtheorem{thm}{Theorem}[section]
\newtheorem{alg}{Algorithm}[section]
\newtheorem{lem}[thm]{Lemma}

\newtheorem{obs}[thm]{Observation}
\def\proof{{\noindent\sl Proof. \rm}}

\begin{document}

\title{\textbf{Point set stratification and Delaunay depth.}
\footnote{M. Abellanas is partially supported by
 \emph{MCYT} TIC2003-08933-C02-01
 M. Claverol and F. Hurtado are partially supported by
\emph{DURSI} 2001SGR00224 and \emph{MCYT} BFM2003-0368}}
\date{}

\author{
\small
Manuel Abellanas $\dagger$, Merc\`e Claverol $\ddagger$, Ferran Hurtado$\S$ \\ \\
\small
$\dagger$Dept. MA, FI UPM, Boadilla del Monte, 28660 Madrid, Spain. mabellanas@fi.upm.es\\
\small
$\ddagger$Dept. MAIV, UPC, Avda. V\'{\i}ctor Balaguer s/n 08800 Vilanova i La Geltr\'{u}, Spain. merce@ma4.upc.edu\\
\small $\S$Dept. MAII, UPC, Jordi Girona 1-3, 08034 Barcelona,
Spain. Ferran.Hurtado@upc.edu}

\maketitle
\thispagestyle{empty}

\begin{abstract}
In the study of depth functions it is important to decide whether
we want such a function to be sensitive to multimodality or not.
In this paper we analyze the Delaunay depth function, which is
sensitive to multimodality and compare this depth with others, as
convex depth and location depth. We study the stratification that
Delaunay depth induces in the point set (layers) and in the whole
plane (levels), and we develop an algorithm for computing the
\emph{Delaunay depth contours}, associated to a point set in the
plane, with running time $O(n\log^2 n)$. The depth of a query
point $p$ with respect to a data set $S$ in the plane is the depth
of $p$ in $S\cup \{p\}$. When $S$ and $p$ are given in the input
the Delaunay depth can be computed in $O(n\log n)$, and we prove
that this value is optimal.
\end{abstract}

\emph{Key words}: {Tukey depth, halfspace depth, convex depth, 
Delaunay depth, depth contours, layers.}

\section{Introduction}

In multivariate analysis classical parametric methodologies are
sensitive to outlying data points and rely on assumptions about
the underlying distribution (as normality or some kind of
symmetry). Data depth has been considered as a measure of how deep
or central a given point is with respect to a multivariate
distribution. Recently nonparametric methods have been developed
based on the concept of data depth \cite{LPS99}. The affine
invariance property of data depth and the spatial ordering of the
sample points leads to the introduction of different methods for
analyzing multivariate distributional characteristics. A survey of
statistical applications of multivariate data depth may be found
in \cite{LPS99}. Several different notions of depth have been
considered, as for instance: \emph{location depth}, also known by
\emph{halfspace depth or Tukey depth} \cite{Tu75}, \emph{convex
depth} or \emph{convex hull peeling depth} \cite{Hu72},
\cite{Ba76}, \emph{Delaunay depth} \cite{Gre81}, \emph{Oja depth}
\cite{Oja83}, \emph{simplicial depth} \cite{Liu90}
and \emph{regression depth} \cite{RH99}.
We can see a classification of multivariate data depths based on their
statistical properties in \cite{ZS00}.

Every notion of depth of a point with respect to a point set $S$
gives rise to a partition of the set $S$ into \emph{layers} and
also to a partition of the whole plane into \emph{levels}. The
layers are the subsets of points of $S$ having the same depth. The
levels are the regions of points in the plane with the same depth
with respect to $S$ (the depth of a  point $p$ with respect to $S$
is the depth of $p$ in $S\cup \{p\}$). The boundaries of the
levels are known by \emph{depth contours} and provide a quick and
informative overview of the shape and some properties of the point
set. For this reason, Tukey suggested the use of depth contours as
a nice tool for data visualization \cite{Tu75}.

Obviously, for any specific purpose of a given statistical
analysis, certain notions of depth may be more suitable than
others. In \cite{OBS92}(pg. 363) Okabe et al. mention the interest
of comparing \emph{Delaunay depth} with respect to other depths.
In this paper we focus on Delaunay depth and compare the
properties of layers and levels associated to finite sets of
points in the plane to the case of \emph{convex depth},
\emph{location depth}. A thorough study is presented in
\cite{Cla04}.

A main concern in current theoretical research on data depth is to
find the \emph{depth contours} and \emph{central regions} by which
the underlying distribution may be characterized. In the discrete
geometry literature, the \emph{center} is any point with location
depth greater than or equal to $\lceil n/(d+1)\rceil$ in
$\mathbb{R}^d$. The center is a point with global maxima depth in
the case of location depth or convex depth and the region of
centers is a connected set; the situation is differently for
Delaunay depth, as shown later, yet it may be desirable to
consider the local maxima keeping in mind the multimodality
features of the underlying set of points.
 Delaunay depth works well on general distributions and is
better than others depths in some respects since it is sensitive
to the existence of clusters and neighborhood relations between
the points. Many interpolation methods are based on Voronoi
diagrams and Delaunay triangulations as a natural neighbor
interpolation method \cite{Sib81}. A selection of clustering
methods is presented in \cite{SHR97}. Different schemes have been
proposed for cluster representation; for example, in \cite{Epps97}
a hierarchical clustering algorithm is developed, and in
\cite{NTM01} another clustering algorithm based on closest pairs
is described.

For every notion of depth, the \emph{median} is defined as a point
with maximal depth. When this point is not unique, the median is
often taken to be the centroid of the deepest region. In
particular, and regarding the applications to statistics, several
medians have been explicitly considered: the Tukey median, the
convex depth median, the maximum simplicial depth median, and the
minimum Oja depth median, as well as a line or a flat with maximum
regression depth. An overview of several multivariate medians and
their basic properties can be found in \cite{Sma90}. The Tukey
median can be used as a point estimator for the data set, and it
is robust against outliers, does not rely on distances, and is
invariant under affine transformations. The location depth and the
corresponding median have good statistical properties as well
\cite{BH99}. Rousseeuw and Struyf present a complete survey about
depth, median, and related measures in \cite{RS04}.

After introducing the basic definitions in
Section~\ref{section:prel}, we give an algorithm in
Section~\ref{section:strat} for computing the \emph{Delaunay depth
contours} (boundaries of the levels), associated to a point set in
the plane. Therefore, we will know the \emph{Delaunay median}
after computing all the levels within the running time of the
algorithm, which is $O(n\log^2 n)$ (where $n$ is the number of
points in the input). We also study and compare the complexity of
the layers and levels of the convex, location and Delaunay depths.
In particular, we see that the depth of a point $p$ with respect
to a set of data $S=\{s_1,\cdots,s_n\}$ can be found in $O(n\log
n)$ time. Lower bounds for this kind of problems have attracted
significant attention, and in Section~\ref{section:computing D
depth} we carry out a study similar to those by Aloupis et al. in
\cite{ACG$^+$02} and \cite{AMcL04}, proving an $\Omega(n\log n)$
lower bound for Delaunay depth computation.

\section{Preliminaries}\label{section:prel}

Let $S$ be a set of $n$ points in the plane, $CH(S)$ the convex
hull of $S$ and $p$ any point of $S$. Any generic depth of $p$
with respect to $S$ is denoted by $d_S(p)$ and the levels and layers of $S$ by
$Lev_i(S)$ and $Lay_i(S)$, respectively. For the specific cases we study we add
superscripts as indicted in the following paragraphs.

The \emph{convex depth} of $p$, is defined recursively as follows:
if $p \in CH(S)$, $d_S^C(p)=1$, else $d_S^C(p)=d^C_{S\setminus
CH(S)}(p)+1$. For values of $j \leq \lfloor n/2 \rfloor$ we say
that the \emph{location depth} of $p$ is  $d_S^L(p)=j$ if and only
if there is a line through $p$ leaving exactly $j-1$ points on one
side, but no line through $p$ separates a smaller subset. The
\emph{Delaunay depth} of $p$, $d_S^D(p)$, is defined to be $d+1$ when the graph
theoretical distance from $p$ to $CH(S)$ in the Delaunay
triangulation $DT(S)$ of $S$ is $d$. In all three cases we call
\emph{depth of} $S$ the depth of its deepest point.

\begin{table*}
 \begin{tabular}{lccccccccc}

     \multicolumn{1}{|c}{} & & \multicolumn{1}{|c}{\small{$d_{S}(p)$ (Depth)}}
    & \multicolumn{1}{|c}{\small{$Lay_i(S)$ (Layer $i$)}}& \multicolumn{2}{|r|}{\small{$Lev_i(S)\quad $(Level $i$)}}
    \\
    \cline{1-6}
     \multicolumn{1}{|c}{} & & \multicolumn{1}{|c}{} & \multicolumn{1}{|c}{}&\multicolumn{2}{|c|}{}
    \\
    \multicolumn{1}{|c}\small{{\it Convex}} & & \multicolumn{1}{|c}{\small{if $p\in CH(S),\, d_{S}(p)=1\,$}} &
    \multicolumn{1}{|c}{}&\multicolumn{2}{|c|}{}
    \\
     \multicolumn{1}{|c}{} & & \multicolumn{1}{|c}{\small{else}} & \multicolumn{1}{|c}{}&\multicolumn{2}{|c|}{}
    \\
\multicolumn{1}{|c}{}& & \multicolumn{1}{|c}{\small{$d_{S}(p)=d_{S\setminus
CH(S)}(p)+1$}} &   \multicolumn{1}{|c}{}&\multicolumn{2}{|c|}{}
    \\
\multicolumn{1}{|c}{}& & \multicolumn{1}{|c}{} &   \multicolumn{1}{|c}{\small{$Lay_i(S)=CH(S_i)$}}&\multicolumn{2}{|c|}{}
    \\
    \cline{1-3}
\multicolumn{1}{|c}{}& & \multicolumn{1}{|c}{} &   \multicolumn{1}{|c}{\small{$S_i=\{x\in S/d_{S}(x)=i\}$}}&\multicolumn{2}{|c|}{\small{Depth of a point}}
    \\
     \multicolumn{1}{|c}\small{{\it Location}} & & \multicolumn{1}{|c}{\small{$d_{S}(p)=j,\quad j\leq \lfloor |S|/2 \rfloor  \Leftrightarrow$} }
    &\multicolumn{1}{|c}{}
    & \multicolumn{2}{|c|}{relative to a set
    $S$}
    \\
     \multicolumn{1}{|c}{\small{}} & & \multicolumn{1}{|c}{\small{some line through $p$
    leaves}}
    &\multicolumn{1}{|c}{\small{}}
    & \multicolumn{2}{|c|}{\small{$d(p,S)=d_{S\cup \{p\}}(p)$}}
    \\
  \multicolumn{1}{|c}{} & & \multicolumn{1}{|c}{\small{exactly $j-1$ points on one}} &\multicolumn{1}{|c}{}
    & \multicolumn{2}{|c|}{\small{}}
    \\
     \multicolumn{1}{|c}{} & & \multicolumn{1}{|c}{\small{side, none leaves less}} &\multicolumn{1}{|c}{}
    & \multicolumn{2}{|c|}{}
    \\
\multicolumn{1}{|c}{}& & \multicolumn{1}{|c}{} &   \multicolumn{1}{|c}{}&\multicolumn{2}{|c|}{}
    \\
    \cline{1-4}
    \multicolumn{1}{|c}{}& & \multicolumn{1}{|c}{} &   \multicolumn{1}{|c}{}&\multicolumn{2}{|c|}{\small{$Lev_i(S)$=}}
    \\
 \multicolumn{1}{|c}\small{{\it Delaunay}} & & \multicolumn{1}{|c}{\small{if $p\in CH(S),\, d_{S}(p)=1\,$}}
    &\multicolumn{1}{|c}{\small{}}
    & \multicolumn{2}{|c|}{\small{$\{x\in
    \mathbb{R}^2/d(x,S)=i\}$} }
    \\
     \multicolumn{1}{|c}{} & & \multicolumn{1}{|c}{\small{else}} & \multicolumn{1}{|c}{$Lay_i(S)=$subgraph of}&\multicolumn{2}{|c|}{}
    \\
    \multicolumn{1}{|c}{} & & \multicolumn{1}{|c}{\small{$d_{S}(p)=$ distance from $p$
     }}
     &\multicolumn{1}{|c}{\small{$DT(S)$ induced by $S_i$}}
    & \multicolumn{2}{|c|}{}
    \\
   \multicolumn{1}{|c}{}  & & \multicolumn{1}{|c}{\small{to $CH(S)$ +1,\,in $DT(S)$}}
     &\multicolumn{1}{|c}{\small{$S_i=\{x\in S |d_S(x)=i\}$}}
    & \multicolumn{2}{|c|}{}
    \\
     \multicolumn{1}{|c}{}  & & \multicolumn{1}{|c}{}
     &\multicolumn{1}{|c}{}
    & \multicolumn{2}{|c|}{}
    \\
    \cline{1-6}
     & & \multicolumn{1}{c}{}
    &\multicolumn{1}{c}{\small{Table $1$: Definitions}}\label{table $1$}
   & \multicolumn{2}{c}{}
  \end{tabular}
\end{table*}

\begin{figure}[ht]
\begin{center}
\includegraphics[width=0.75\columnwidth] {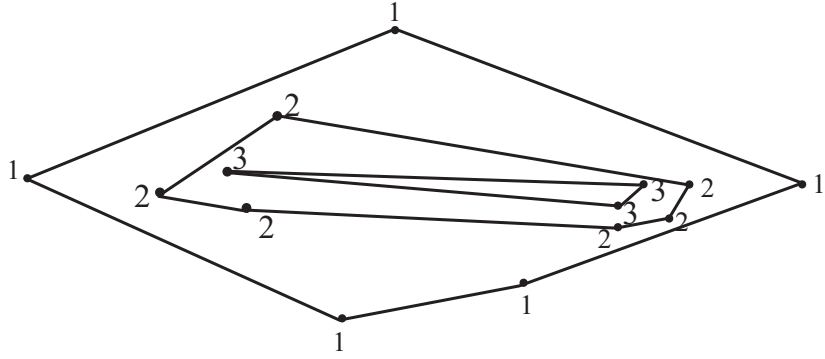}
\caption{Convex layers.}\label{figcapasc}
\end{center}
\end{figure}

\begin{figure}[ht]
\begin{center}
\includegraphics[width=0.75\columnwidth] {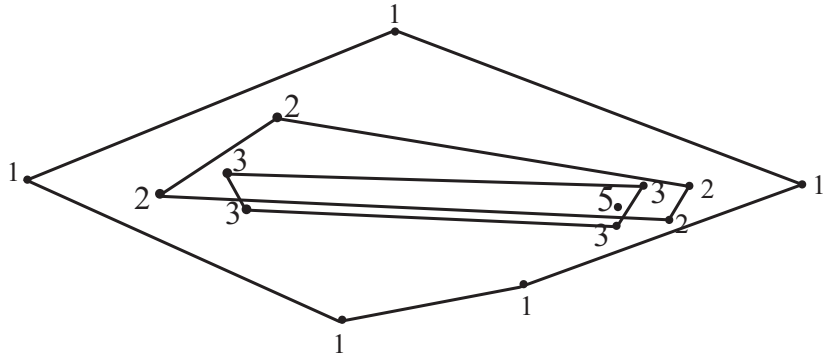}
\caption{Location layers.}\label{figcapasl}
\end{center}
\end{figure}

\begin{figure}[ht]
\begin{center}
\includegraphics[width=0.75\columnwidth] {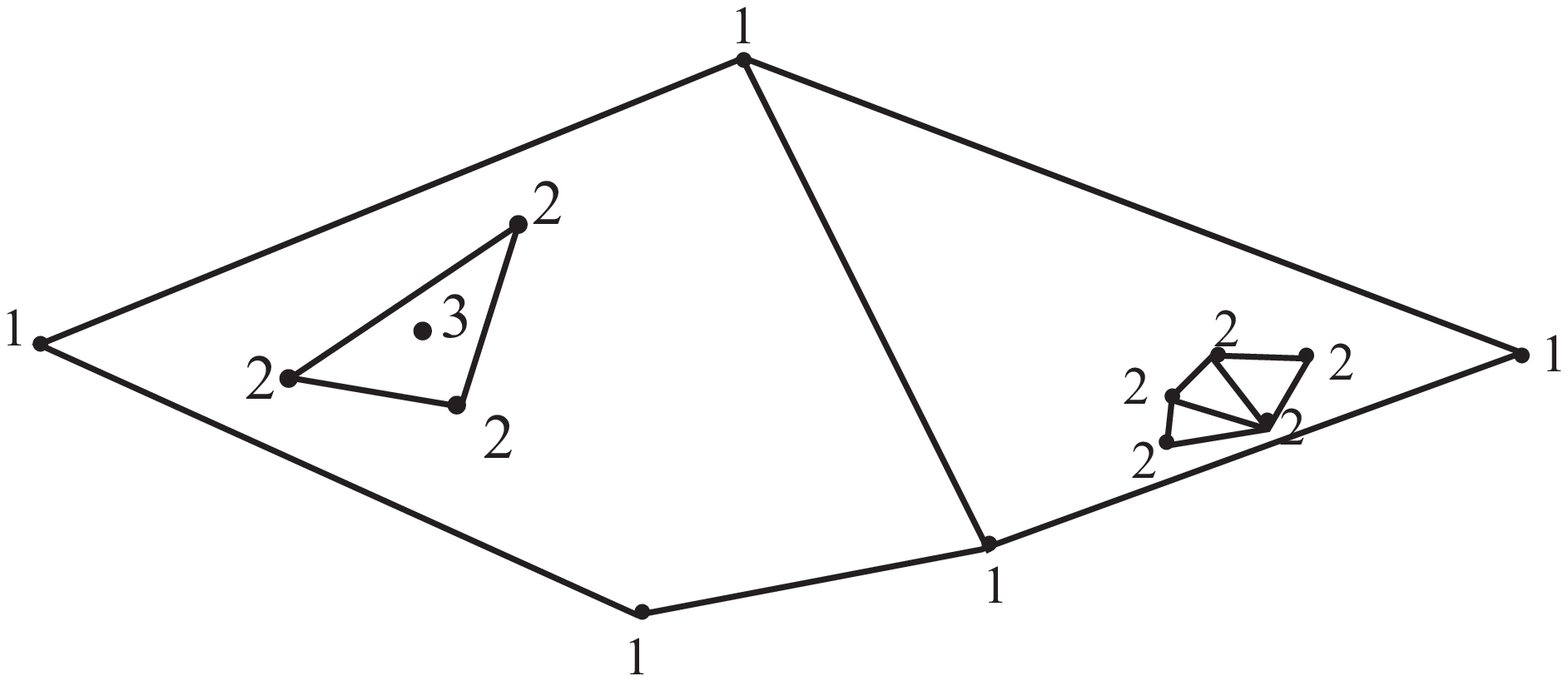}
\caption{Delaunay layers.}\label{figcapasd}
\end{center}
\end{figure}

The  $i$-th \emph{layer} of $S$, $Lay_i(S)$, is defined for convex
depth as well as for location depth by $Lay^C_i(S)=Lay^L_i(S)=CH(S_i)$, where
$S_i=\{x\in S\; |\;d_S(x)=i\}$, (Figures \ref{figcapasc} and \ref{figcapasl}).
For the Delaunay depth, $Lay^D_i(S)$
is the subgraph of $DT(S)$ induced by $S_i$, (Figure \ref{figcapasd}).

Let $p$ be any point in the plane. For the three depths
considered, the depth of $p$ relative to the set $S$ is
$d(p,S)=d_{S\cup \{p\}}(p)$ and the $i$-th \emph{level}  for the
set $S$ is defined by $Lev_i(S)=\{x\in \mathbb{R}^2| d(x,S)=i\}$.
The concept of $k$-hull introduced by Cole, Sharir and Yap in
~\cite{CSY87} corresponds to $\bigcup_{j\geq k}Lev_j(S)$,
also know by kth depth region $D_k$.

Table 1 shows all these definitions together.

\section{Point set stratification}\label{section:strat}

\begin{figure}[ht]
\begin{center}
\includegraphics[width=0.75\columnwidth] {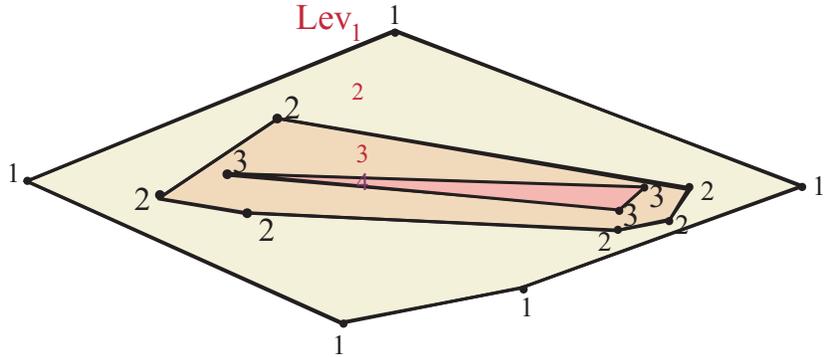}
\caption{Convex levels.}\label{fig.clevels}
\end{center}
\end{figure}

Given a set $S$ of $n$ points in the plane the convex layers can
be constructed with Chazelle's optimal $O(n\log n)$ algorithm
\cite{Cha85}. Convex layers form a sequence of nested convex polygons
defining a partition of the plane into regions, which coincide
with the levels, (Figures \ref{figcapasc} and \ref{fig.clevels}).
Therefore layers and levels have linear
complexity in the convex depth case and can be constructed in
optimal $O(n\log n)$ time.

\begin{figure}[ht]
\begin{center}
\includegraphics[width=0.75\columnwidth] {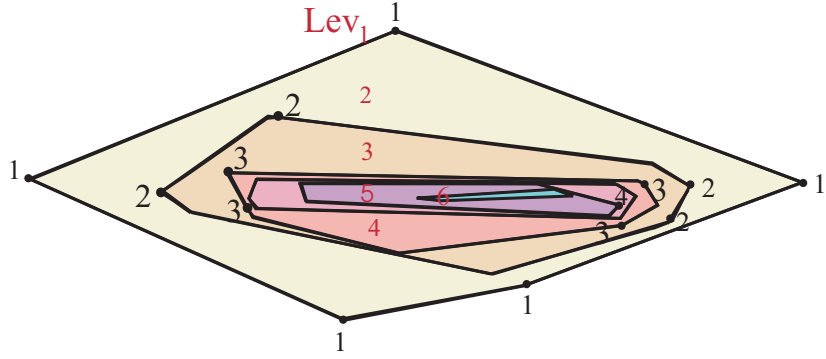}
\caption{Location levels.}\label{fig.loclevels}
\end{center}
\end{figure}

As for location depth, a worst case optimal algorithm for
computing all $Lev_i^S(S)$, (where $n/3\leq i\leq n/2$) in $O(n^2)$
time is obtained by using topological sweep in the dual arrangement
of lines (see \cite{Cla04}, \cite{MRR$^+$03}). The boundaries of the
levels, in this case, form a sequence of nested convex polygons. Points of
$Lay_i^S(S)$ are in convex position and belong to the boundary of
$Lev_i^S(S)$, but this boundary can also have other vertices not in
$S$, (Figure \ref{fig.loclevels}). Some layers can be empty and
different layers can cross each other (Figure \ref{figcapasl}).
While the complexity of levels may reach $O(n^2)$, the size
of the layers is $O(n)$. The layers in the location depth case can
be computed using the mentioned $O(n^2)$ sweep algorithm yet, to
our knowledge, it is an open problem to construct them in less
time or to prove a quadratic lower bound for the problem.

\begin{figure}[ht]
\begin{center}
\includegraphics[width=0.75\columnwidth] {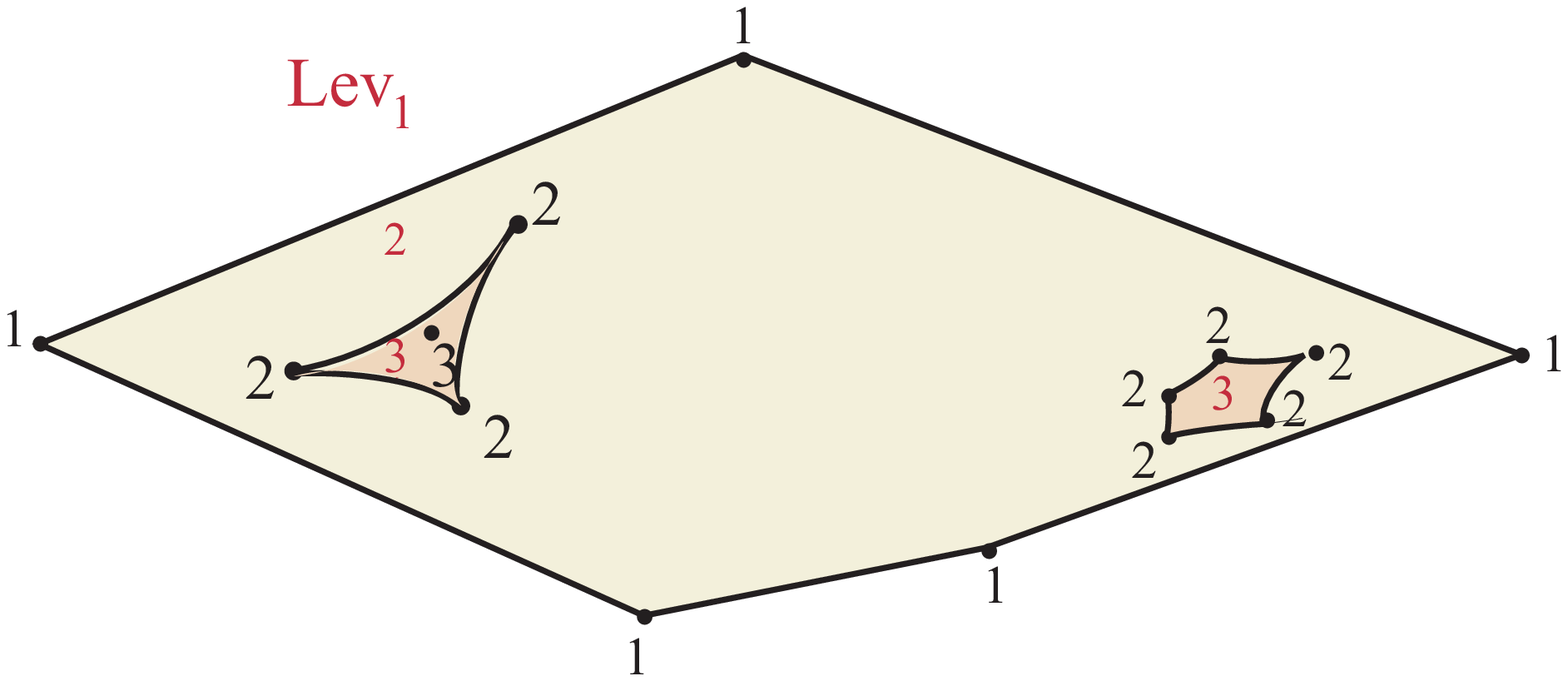}
\caption{Delaunay levels.}\label{fig.dlevels}
\end{center}
\end{figure}

Much less has been studied to Delaunay depth, which we explore
sistematically in the rest of this section.

In the Delaunay depth case, all the layers $Lay_i^D(S)$, $i\leq n/3$, can
easily be found by visiting $DT(S)$ in linear time once
constructed, which requires $O(n\log n)$ time (Figure
\ref{figcapasd}). Notice that one layer can have more than one
connected component.
Next, we study the Delaunay layers. First, we show some properties of
Delaunay layers which allow us to obtain the levels easily and also to
prove other results as that the $\bigcup Lev_i^D(S)$ are nested sets.
Next, we will study the number of connected components that we can have
in the $\bigcup Lay_i^D(S)$.

\begin{prop}\label{pesomayor}
Let $S$ be a set of Delaunay depth greater than one.
The points of $S$, in the interior of any cycle $C_i$ of $Lay_i^D(S)$,
have depth greater than $i$.
\end{prop}

\proof
Let $p\in S$ be, which is in the interior of a cycle $C_i$ of $Lay_i^D(S)$.
From the definition of Delaunay depth, we know that $p$ must have some
adjacency of depth $d_S^D(p)-1$.
The points adjacent to $p$ are points of $C_i$ or they are in the interior of $C_i$.

If we suppose the assertion of the proposition is false, $d_S^D(p)\leq i$.
Then there exists a point $q$ adjacent to $p$, with $d_S^D(q)=d_S^D(p)-1$
and interior of $C_i$.
Recursively it follows that there is at least a point of depth equal to $1$ in
the interior of $C_i$, which is impossible. Then we conclude that all points
of $S$ which are in the interior of $C_i$ have depth greater than $i$.

$\hfill \Box$

\begin{lem} \label{unacc}
Let $S$ be a set of Delaunay depth greater than one.
Any cycle of $Lay_i^D(S)$ without chords, does not contain more than one
connected components of $Lay_{i+1}^D(S)$ in its interior.
\end{lem}

\proof
Let $C_i$ be a cycle of $Lay_i^D(S)$ formed by points without chords.

Suppose, contrary to our claim, that there are more than one connected
component of $Lay_{i+1}^D(S)$ in the interior of $C_i$.
By the above assumption, we first prove that there is a vertice $v_i\in C_i$
which is adjacent to some points of different connected components of
$Lay_{i+1}^D(S)$ in the interior of $C_i$ (Figure~\ref{fig6.12}).
Let $v_i^1,v_i^2,\cdots ,v_i^n$ be the points of $C_i$ sorted by adjacencies.
We study the adjacencies of these points in the interior of $C_i$.
Note that this adjacencies have depth equal to $i+1$ (we apply that their depth
cannot differ more than one of $i$ and Proposition \ref{pesomayor}); furthermore,
all the points of $Lay_{i+1}^D(S)$ in the interior of $C_i$ must have at least
one adjacency in $C_i$.

\begin{figure}[ht]
\begin{center}
\includegraphics[width=0.3\columnwidth]{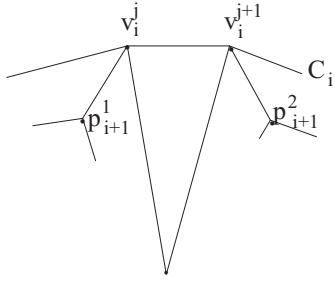}
\caption{The points $v_i^j$ and $v_i^{j+1}$ of $C_i$ have adjacencies of different
components of $Lay_{i+1}^D(S)$.}\label{fig6.11}
\end{center}
\end{figure}

\begin{figure}[ht]
\begin{center}
\includegraphics[width=0.3\columnwidth]{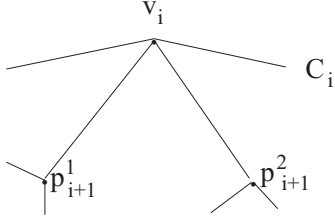}
\caption{There is a point $v_i\in C_i$ adjacent to some points of different
connected component of $Lay_{i+1}^D(S)$ in the interior of $C_i$.}\label{fig6.12}
\end{center}
\end{figure}

We move along $C_i$ following the adjacencies:
while the adjacencies are of the same connected component we are changing of
point in $C_i$. We want to find different connected components in the adjacencies.
There are two possibilities:
\begin{enumerate}
\item There is a point $v_i^j$ which is adjacent to some points of different connected
components of $Lay_{i+1}^D(S)$ in the interior of $C_i$.
\item There are $v_i^j$ and $v_i^{j+1}$, for some $j$, whose adjacencies are in
different components of $Lay_{i+1}^D(S)$ (Figure~\ref{fig6.11}).
\end{enumerate}

But in the second case, we can see that the point $v_i^j$ or $v_i^{j+1}$ must
also have adjacencies in different components of $Lay_i(S)$ (is a point like in the first case).
In order to prove that, we can consider the point which forms a triangle in
the $DT(S)$ with $v_i^j$ and $v_i^{j+1}$. This point can only be of depth $i+1$;
it cannot be $i$ because then $v_i^jv$ and $v_i^{j+1}v$ would be chords, contrary
of the hypothesis of the proposition. Hence, $v\in Lay_{i+1}^D(S)$ but, $v$ cannot
belong at the same time to the different connected components where $v_i^j$
and $v_i^{j+1}$ have adjacencies.

We have proved that $v_i\in C_i$ exists with two adjacencies of different components
of $Lay_{i+1}^D(S)$, we denote them by
$p_{i+1}^1,p_{i+1}^2$ like Figure~\ref{fig6.12}.
Then there is a path in the $DT(S)$ between $p_{i+1}^1$ and $p_{i+1}^2$ formed by
a sequence of vertices of triangles which all they have $v_i$ as point in common.
Note that this sequence only can be formed by points of depth $i+1$:
there is no point with depth $i+2$ because this point is adjacent to $v_i$, of depth
$i$ and also there is no a point of depth equal to $i$ because this point with $v_i$
would be a chord which contradicts the assumptions.
The $Lay_{i+1}^D(S)$ is formed by the subgraph induced in the $DT(S)$ by the points
with the same depth, so all the points adjacents to $v_i$, between $p_{i+1}^1$ and
$p_{i+1}^2$, are in the same connected component, a contradiction.

Hence we conclude that any cycle of $Lay_i^D(S)$ without chords, does not contain
more than one connected component of $Lay_{i+1}^D(S)$ in its interior.

$\hfill \Box$

\begin{lem}\label{lem.existscycle}
Let $S$ be a set of points in the plane. Let $p\in Lay_{i+2}^D(S)$ and
let $C^i$ be a cycle of $Lay_i^D(S)$ that contains $p$ in its interior.
Then there is a cycle of $Lay_{i+1}^D(S)$ containing $p$ in its interior.
\end{lem}

\proof
Let $p_{i+2}\in$ D-$Lay_{i+2}(S)$ be a point in the interior of $C_i$.
From Lema \ref{unacc} we know that there is only one connected component of D-$Lay_{i+1}(S)$ in $C_i$.

When we consider a connected graph without cycles embedded in the plane, there is only a single infinite region, complementary to the graph. If the graph has some cycles, then we distinguish the bounded regions enclosed by the edges of the cycles. We will prove that the graph $G$ formed by the points with depth $i+1$ inside $C_i$ must be a graph with cycles. Its unbounded region contains $C_i$.
Each point of the considered graph $G$ has depth $i+1$ and it is adjacent to one of the $C_i$.
We consider the Delaunay triangles with at least one vertex in $C_i$. The point $p_{i+2}$ cannot be vertex of any of those triangles (the depths cannot differ in more than one unit).
The union of those triangles does not contain $p_{i+2}$ because the Delaunay triangles do not contain points of $S$ in their interior.
Only if $G$ has some cycles, there can be other points placed in the bounded regions delimited by them.
Therefore, if there exists a point of D-$Lay_{i+2}(S)$ in the interior of $C_i$, then there exists too a cycle of D-$Lay_{i+1}(S)$ containing such point in its interior.

$\hfill \Box$

\begin{prop}\label{prop6.6}
Let $S$ be a set of points in the plane. If the Delaunay depth of a point
$p$ with respect to $\,S\,$ is $j+1$, there is a cycle of $Lay_{j}^D(S)$
containing $p$ in its interior.
\end{prop}

\proof
Every point $p$ whose depth with respect to $\,S\,$ equals $2$, is contained in
the interior of $Lay_1^D(S)=CH(S)$.

If the depth of $p$ is $3$, there exists a cycle of $Lay_2^D(S)$ containing $p$
in its interior. In order to prove that, we apply Lemma \ref{lem.existscycle} to
a cycle of points of depth $1$ that contains $p$ (this cycle exists because
$Lay_1^D(S)=CH(S)$).

If the depth of $p$ is $4$, there is a point of $Lay_3^D(S)$ adjacent to $p$.
We apply Lemma \ref{lem.existscycle} to this point of depth $3$.
Then there is a cycle of $Lay_2^D(S)$ that contains this point of depth $3$,
and must contain its adjacencies, like $p$. We apply lemma \ref{lem.existscycle}
to this last cycle and there is a cycle of $Lay_3^D(S)$ that contains $p$.

Recursively we prove the proposition for $p$ of depth $j+1$
$\forall j, j\leq f-1$ ($f$ being the depth of $S$).

$\hfill \Box$

As a consequence of Proposition \ref{prop6.6}, the number of levels
for Delaunay depth is equal to the number of layers or to the number of layers plus one.

\begin{prop} \label{ccenmcapas}
Let $S$ be a set of $n$ points. The maximum number of connected
components of the $\bigcup Lay_i^D(S)$ is decreasing on the depth
of $S$. This maximum is $\displaystyle{ \lfloor (n-m+2)/2\rfloor
}$, where $m$ is the depth of $S$, which is tight.
\end{prop}

\proof
We want to see that $c$, the number of connected components of
$\bigcup Lay_i^D(S)$, is bounded by $(n-m+2)/2$ or, equivalently,
$n\geq 2c+m-2$.

If all the related connected components have a minimum of $2$
points, then $n\geq 2c$. If there are isolated points in
$Lay_{i+1}^D(S)$, each one of them is contained in a cycle without
chords (Proposition \ref{lem.existscycle}). We associate each
isolated point with a point of the corresponding cycle in this
way: two isolated points cannot be associated to the same point.
This is possible because the maximum number of the isolated points
of $Lay_{i+1}^D(S)$, contained in a connected component of
$Lay_i^D(S)$, is at most the number of chords plus one (Lemma
\ref{unacc}). Moreover, the number of chords in a connected
component of $n_i$ points is at most $n_i-3$ so there are no
points of depth $i$ in the interior of a cycle of $Lay_i^D(S)$
(Proposition \ref{pesomayor}).

Then, there are at least two points in each component that are not
associated to any of the possible isolated points. Thus we can
assure $n\geq 2c$.

In general, if the depth of $S$ is $m$, there exist at least $m-1$
nested cycles, without chords, of which $m-2$ don't contain any
component of a single point. The connected component that contains
one of the previous cycles have, at most, $n_i-3$ isolated points.
Therefore, there are at least $m-2$ connected components with
three points or more. Then $n\geq 2c+m-2$.

The next example proves that the previous upper bound is tight.

\begin{figure}[ht]
\begin{center}
\includegraphics[width=0.5\columnwidth]{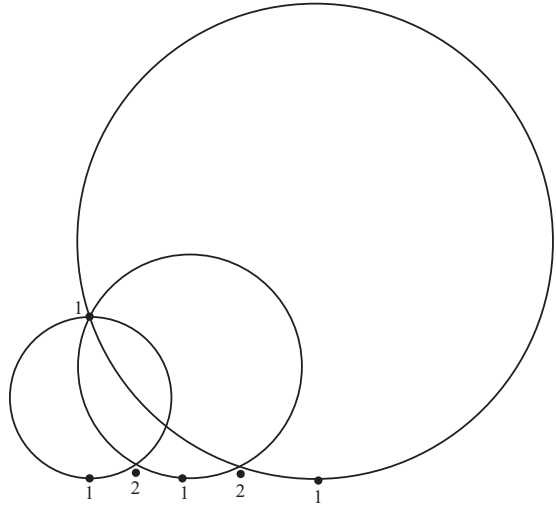}
 \caption{This is a example in which $n=2k+2$ points. The $\bigcup Lay_i^D(S)$
 has $k+1$ connected components.}\label{fig6.10}
\end{center}
\end{figure}

First we describe the example for $m=2$.
Let $n=2k+2$ be the number of points that we have. We distinguish two chains in
the $CH(S)$: in one of them (for example the lower chain) we put $k+1$ of the
points of $S$ and in the other (the upper chain) we put only one point.
We can place the points in this way: for every pair of points formed with the upper chain point and
any lower chain point, there must be an empty circle that circumscribes them.
Finally, we put
each one of the other $k$ points of $S$ between two of the previous circles like
in Figure~\ref{fig6.10}. These $k$ points are each one of them one connected component
of $Lay_2^D(S)$, so the $\bigcup Lay_i^D(S)$ has $k+1=\lfloor n/2\rfloor$ connected components.

\begin{figure}[ht]
\begin{center}
\includegraphics[width=0.8\columnwidth]{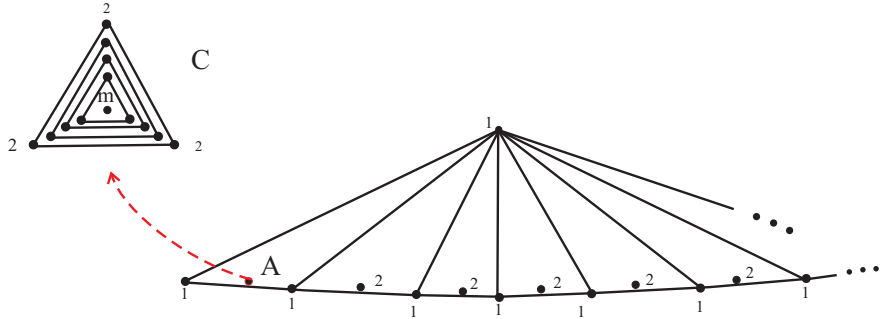}
 \caption{Point $A$ is replaced by configuration $C$. The set of points $S$ has depth equal to $m$. The $\bigcup Lay_i^D(S)$ has $\lfloor (n-m+2)/2\rfloor$ connected components.}\label{fig6.10.1}
\end{center}
\end{figure}

Let $m$ be greater than $2$. First we put $3(m-1)$ points in a
sequence of nested triangles and one more point in the innest one.
The rest of the points of $S$, at most $n-3m+2$, are distributed
in pairs between the $m$ layers. We place each pair of the points
in contiguous layers so one of them breaks a cycle in two and the
other one is an isolated point in the new cycle. In figure
\ref{fig6.10.1}, the $n-3m+2$ points have been placed in the
layers $Lay_1^D(S)$ and $Lay_2^D(S)$.

$\hfill \Box$

Delaunay layers are not necessarily polygons, however they form a
structure based in nested cycles of points of the same depth.

The depth of a point relative to a set $S$ depends on the Delaunay
circles (i.e., circumcircles of Delaunay triangles) that contain
the point, therefore the arrangement of Delaunay circles contains
all the information about Delaunay levels,
(Figure~\ref{fig.dlevels}). As the arrangement has size $O(n^2)$
and can be constructed in $O(n^2 \log n)$ time one can obtain the
Delaunay levels within this time. Nevertheless, in the following
theorem we prove that in order to obtain all  $Lev_i^D(S)$ it is
not necessary to construct the whole arrangement of circles.

\begin{obs}\label{observation}
Let $C$ be a circle having exactly two points $u$ and $v$ of
 $S$ on its boundary and containing no points of $S$ in its
interior. Then any circle crossing the two arcs determined by $u$
and $v$ in the boundary of $C$ contains some interior point from
$S$.
\end{obs}

\begin{thm}\label{thm1}
Let $S$ be a set of points in the plane and let be $f$ its
Delaunay depth. The union $\bigcup_{j\geq k}Lev_j^D(S)$,
$k=1,\cdots ,f$ forms a sequence of sets nested by inclusion. The
boundaries between $Lev_j^D(S)$ and $Lev_{j+1}^D(S)$, for $2\leq
j\leq f$, are curves composed by arcs of the Delaunay circles
determined by two points $u, v$ of $Lay_j^D(S)$ and one point $w$
of $Lay_{j-1}^D(S)$.
\end{thm}

\begin{proof}
We proceed to determine the boundary between the consecutive
levels of $S$, $Lev_j^D(S)=\{x\in\mathbb{R}^2/d(x,S)=j\}$, and
$Lev_{j+1}^D(S)$, for $2\leq j\leq f$. Every point $q$ of depth
equal to $j$, relative to a set $S$, has at least one element
$p\in S$ which is adjacent in $DT(S\cup\{q\})$ and has depth $j-1$
(in both $DT(S)$ and $DT(S\cup\{q\})$), and there must be an empty
circle through $p$ and $q$ and no point of $S$ with depth smaller
than $j-1$. Hence we can describe the $Lev_j^D(S)$ as the union of
all Delaunay circles that circumscribe a point of depth $j-1$
(that we denote by $\bigcup C_{j-1,-,-}$), minus the union of all
Delaunay circles that circumscribe a point of depth smaller than
$j-1$ (that we denote by $\bigcup C_{<j-1,-,-}$); this is

$$Lev_j^D(S)=\bigcup C_{j-1,-,-}\setminus \bigcup C_{<j-1,-,-}.$$

Applying Proposition \ref{lem.existscycle}, which proves that for
every point of depth equal to $j$ there is a cycle of
$Lay_{j-1}^D(S)$ that contains it in its interior, we see that
$Lev_j^D(S)$ is contained in the interior of the cycles of
$Lay_{j-1}^D(S)$. Furthermore we also get the following
properties: (a) If some layer has no cycles then there are no
points for this level or the next ones; (b) the sets
$\bigcup_{j\geq k} Lev_j^D(S), \, k=1,\cdots ,f$ form a sequence
of nested sets.

We find circles $C_{j,j,j-1}\in \bigcup C_{<j,-,-}$
intersecting the cycles of $Lay_{j}^D(S)$. The circles
$C_{j,j,j-1}$ pass through pairs of points which are the
endpoints  of every non-chord edge of a cycle $\gamma$ in
$Lay_{j}^D(S)$ (see the cycle of $Lay_3^D(S)$ enclosing the dark
region to the left of Figure \ref{fig:LevDel}).

\begin{figure}[ht]
\begin{center}
 \includegraphics[width=1\columnwidth]{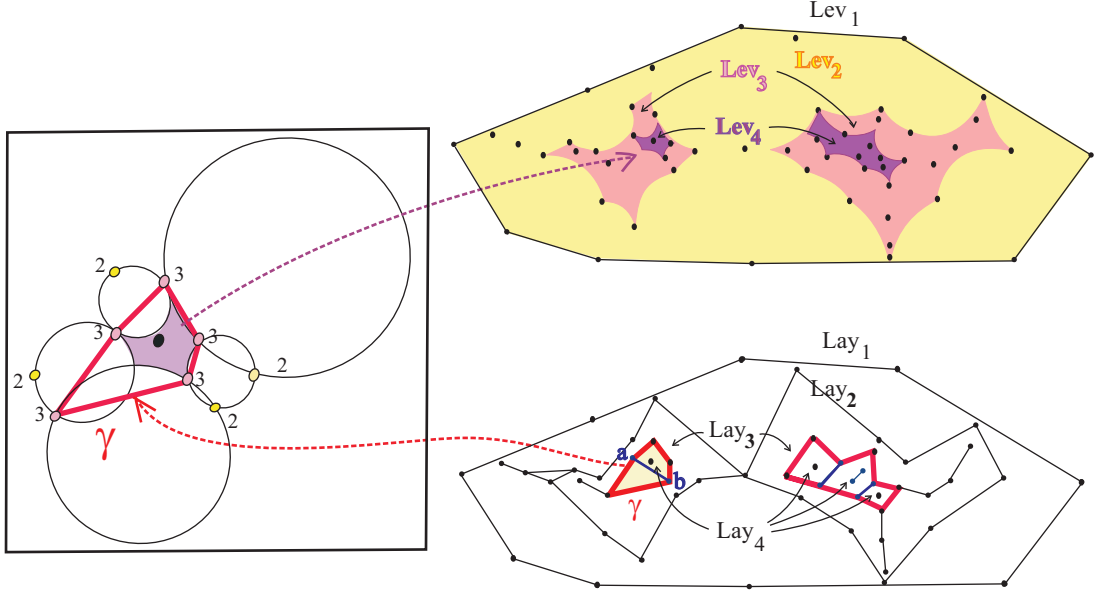} 
\caption{The Delaunay circles
 $C_{3,3,2}$ defined by two points of
$Lay_3^D(S)$ and one point of $Lay_2^D(S)$, determine the
boundary between $Lev_3^D(S)$ and $Lev_4^D(S)$, which consists of
the inner boundary of the union of $C_{3,3,2}$.
Notice that chord $ab$ has been \lq\lq discarded\rq\rq , as
unuseful for obtaing the level.}\label{fig:LevDel}
\end{center}
\end{figure}

These pairs of points divide the circle $C_{j,j,j-1}$ into two
arcs: one exterior to the cycle $\gamma$, one interior. There may
be other circles of $\bigcup C_{<j,-,-}$ that also cross the
circle $C_{j,j,j-1}$, yet any circle of $\bigcup C_{<j,-,-}$
has in the boundary one point exterior to the cycle $\gamma$ and,
applying Observation \ref{observation}, it cannot cross both arcs
of a circle $C_{j,j,j-1}$.

Therefore the boundary between $Lev_{j}^D(S)$ and $Lev_{j+1}^D(S)$
is only determined by the arcs of the circles $C_{j,j,j-1}$
(see Figure \ref{fig:Levj+1} for an illustration).

\begin{figure}[ht]
\begin{center}
\includegraphics[width=0.45\columnwidth] {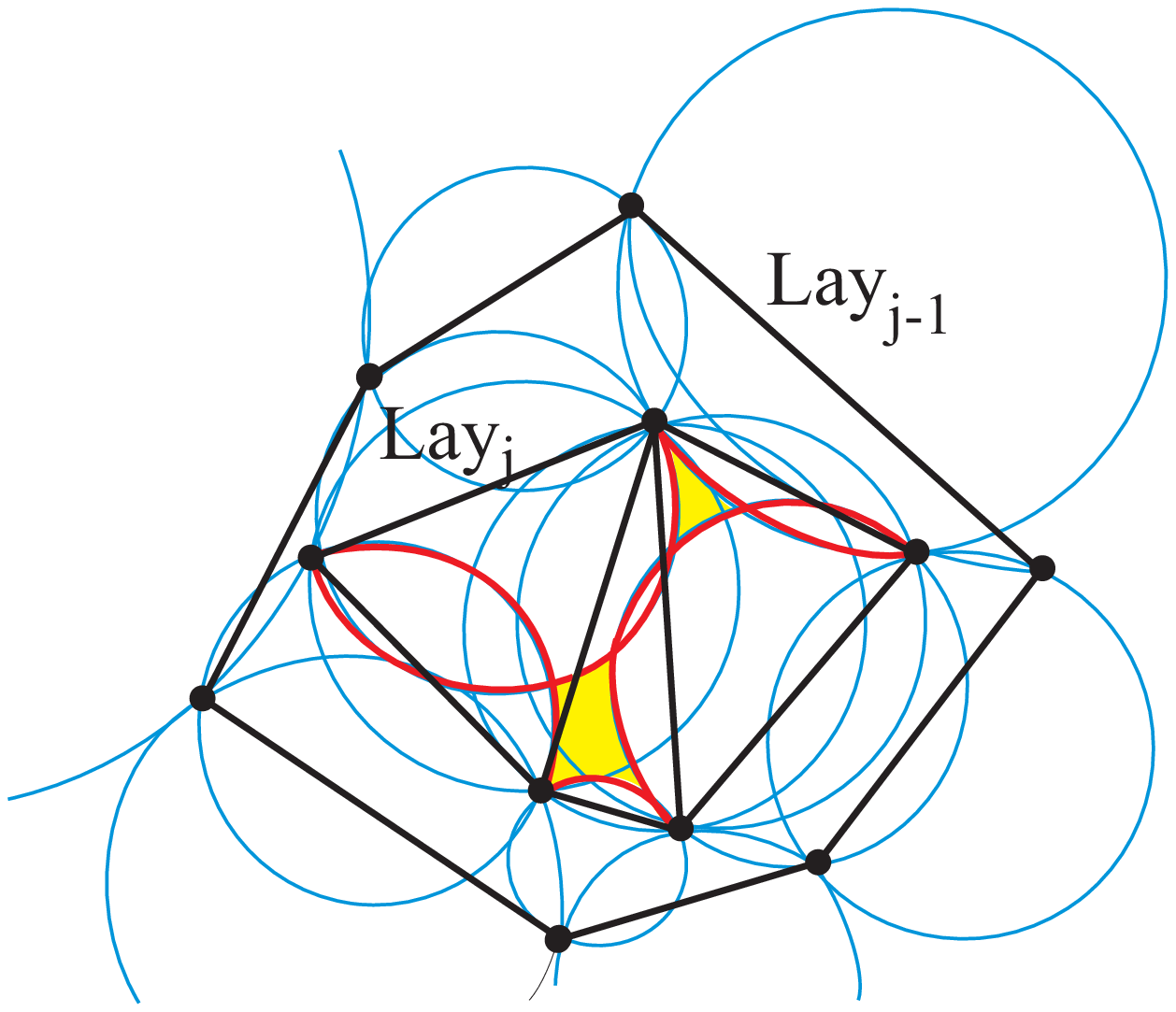} \caption{The
shaded region is $Lev_{j+1}^D(S)$.}\label{fig:Levj+1}
\end{center}
\end{figure}

$\hfill \Box$
\end{proof}

Theorem \ref{thm1} proves that the overall size of the Delaunay
levels is $O(n)$ and justifies the steps of the following
algorithm.

\begin{alg}\label{alg.levels}
\textsc{Computation of Delaunay depth contours of $S$, Delaunay levels.}

\noindent{\sc{Input:}} Set of points $S$.

\noindent{\sc{Output:}} Delaunay depth contours of $S$.
\begin{enumerate}
\item[1.] Compute $DT(S)$.
\item[2.] Compute the Delaunay depths for all points in $S$.
\item[3.] Compute the boundaries of the levels
as follows: $Lev_1^D(S)$ is the convex hull of $S$; for every
$j\geq 2$, construct the inner boundary of the union of Delaunay
circles $C_{j,j,j-1}$ defined by two points $u, v$ of $Lay_j^D(S)$
and one point $w$ of $Lay_{j-1}^D(S)$ (Figure~\ref{fig:Levj+1}).
\end{enumerate}
\end{alg}

$DT(S)$ can be computed in $O(n \log n)$ time and Step $2$ takes
$O(n)$ additional time. Every boundary in Step $3$ can be computed
in $O(t \log^2 t)$ time, where $t$ is the number of Delaunay
circles $C_{j,j,j-1}$ considered in the currently computed layer,
by using the algorithm described in \cite{AS00} (pg. 97). Taking
into account that the total number of Delaunay circles is $O(n)$,
Step $3$ takes $O(n\log^2 n)$ global time, which is also the
overall time for the algorithm. Notice that the expected time for
Step 3 is $O(n \log n)$ \cite{AS00}, and therefore, the expected
running time for the entire algorithm is $O(n \log n)$.

\begin{figure}[ht]
\begin{center}
\includegraphics[width=0.7\columnwidth]{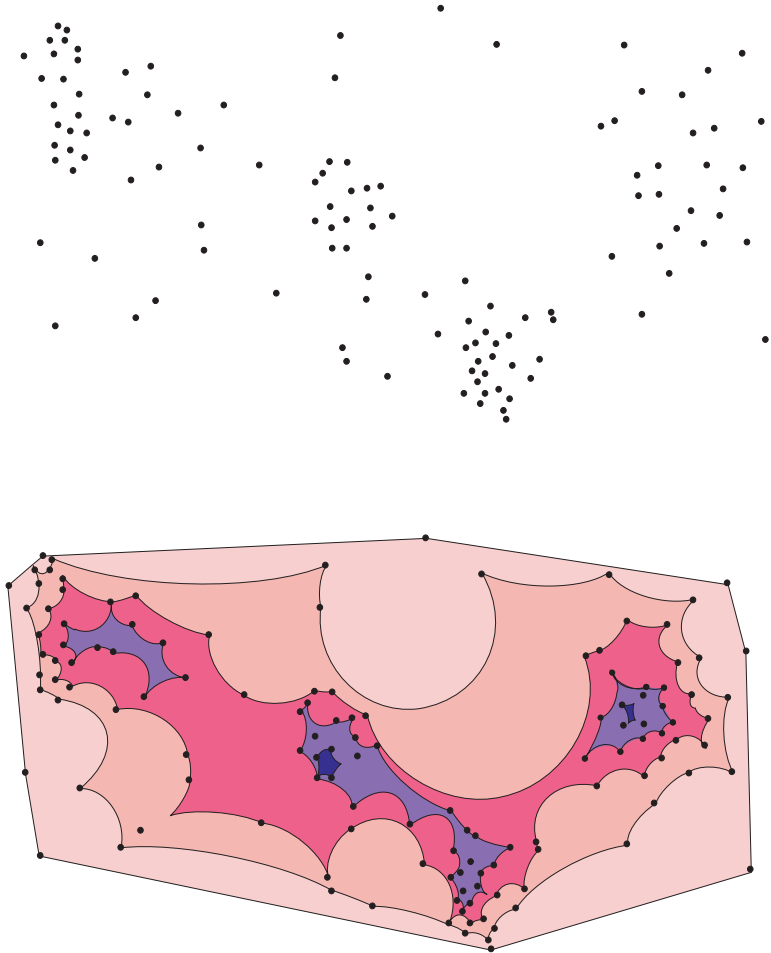}
 \caption{Top: A point set $S$. Bottom: Levels of $S$.
 The boundaries of the levels are the \emph{Delaunay depth contours}.}\label{fig.Dcontours}
\end{center}
\end{figure}

The algorithm \ref{alg.levels} compute all levels of $S$ in
$O(n\log ^2 n)$ time, therefore it also yields the \emph{Delaunay
median} in this time. In Figure \ref{fig.Dcontours} we can see an
illustration where the inner level, $Lev_6^D(S)$, has two
connected components: the centroids of each one of these regions
are the Delaunay median of $S$.

As a consequence of the preceding paragraphs we can state the following theorem.

\begin{thm}\label{thm2}
The Delaunay levels of a set of $n$ points in the plane can ce
constructed within $O(n\log ^2 n)$ time.
\end{thm}

\section{Computing Delaunay depth}\label{section:computing D depth}
The depth of a point $p$ with respect to a data set
$S=\{s_1,\cdots,s_n\}$ in the plane is defined as the depth of $p$
in $S\cup \{p\}$, and its computation is a problem which has
deserved much attention. When $S$ and $p$ are the entry data, the
Tukey depth of $p$, its simplicial depth and its Oja depth can be
computed in $O(n\log n)$ \cite{RR96}. In \cite{ACG$^+$02} it was
proved that this value is also a tight bound for the first two
cases and recently it has been proved an identical result for the
Oja depth \cite{AMcL04} .

The convex depth of $p$ can be easily computed in $O(n\log n)$
time, since it suffices to find the layers of $S\cup \{p\}$, and
it is easy to see that this value is tight. The Delaunay depth can
also be found in $O(n\log n)$, since it suffices to build
$DT(S\cup\{p\})$ and then find the depth of $p$ in additional
$O(n)$ time. We will next show that this is tight.

We will reduce the problem of uniqueness of numbers to the problem
of finding the Delaunay depth.  It is known that the problem of
deciding if, given $n$ real numbers, all of them are distinct, has
complexity $\Omega(n\log n)$ when the model of computation is the
algebraic decision tree \cite{DL76} and \cite{BO83}. We will see
that if certain computations are made in $O(n)$ and then the
Delaunay depth of an adequate point is found, we can decide the
uniqueness of $n$ given real numbers. This implies that the
computation of the Delaunay depth requires $\Omega(n \log n)$
time.

\begin{figure}[ht]
\begin{center}
\includegraphics[width=0.45\columnwidth] {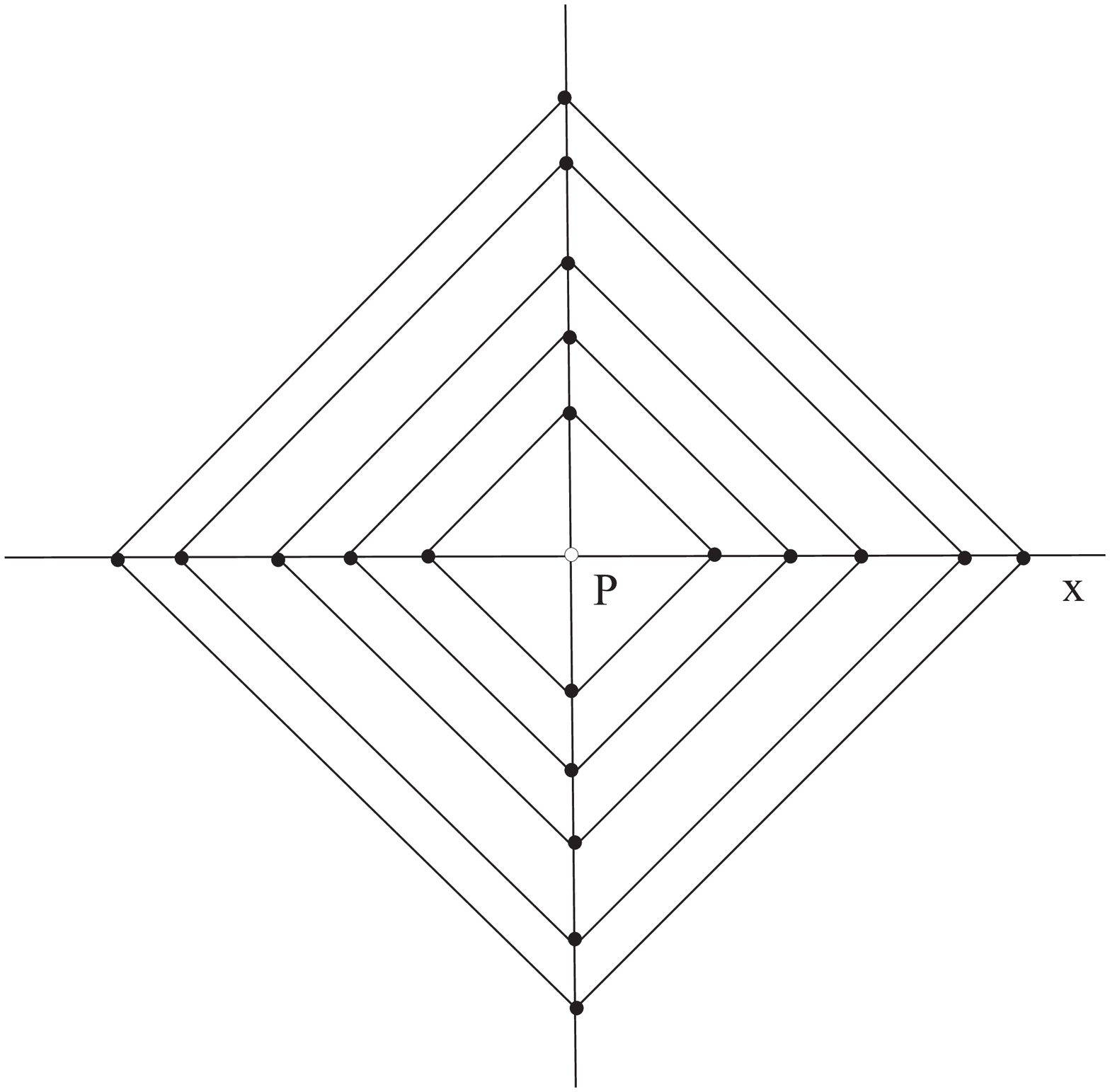}
\caption{Set of points $A$ and its Delaunay layers.}\label{uniqueness}
\end{center}
\end{figure}
Let us consider a set $A=\{x_1,\cdots,x_n\}$ of real numbers;
without loss of generality we can assume that they are all
positive. For each value $x_i\in A$, we construct the points
$(x_i,0),(-x_i,0),(0,x_i)$ and $(0,-x_i)$. We denote by $S$ the
union of these points and let $p=(0,0)$ be the origin. The
Delaunay triangulation $DT(S\cup\{p\})$ is as  shown in Figure
\ref{uniqueness}, from which we have omitted the diagonals of the
trapezium (any of the two diagonals in a trapezium gives a
Delaunay triangulation and the depths of the points remain
unaltered by the choice). The presence of the edges of slopes $\pm
1$ is immediate: for example, $(x_i,0)$ is adjacent to $(0,x_i)$
since the circle of center $(x_i,x_i)$ and radius $x_i$ covers
only these two points of $S\cup \{p\}$.

Evidently, the depth of $p$ in $S\cup \{p\}$ equals $n+1$ if, and only if,
all the elements of $A$ are distinct. This completes the proof. It has thus
been established the following result:

\begin{thm}
The depth of a point $p$ with respect to a data set $S=\{s_1,\cdots,s_n\}$
can be found in $O(n\log n)$ time, and this value is optimal.
\end{thm}

If we admit an additional preprocess to the given point set, we
have different alternatives for computing the level of a new
point. For example the preprocessing might consist of computing
the Delaunay triangulation, or even the arrangement of the
Delaunay circles; nevertheless the most natural approach is to
compute the Delaunay levels in a first step, which requires
$O(n\log^2 n)$ time; as this gives a plane subdivision of size
$O(n)$, standard point-location methods can then be used. In
particular, the approach in \cite{ST86} can be easily adapted and
allows $O(\log n)$ query time.

It is also natural to consider how strong the change in the
Delaunay depths of a point set can be after the insertion of a new
point. This is the issue we study next.

\begin{prop}
Let $S$ be a set of $n$ points of depth equal to $f$. The insertion of one point
in $S$ can change the depth of another point in at most $\lfloor n/3\rfloor -2$
units and the depth of the set can vary by $\displaystyle{\lfloor n/3\rfloor -3}$.
These bounds are tight.
\end{prop}
\begin{proof}
One point can vary its depth when its set of neighbors varies (for instance when $p$ is a new neighbor)
or some of its neighbors changes its depth. The insertion of one point
in $S$ can produce at most a change of depth equal to $f-2$
units, if and only if some of the deepest points is a neighbor of the least deep one.

Let us see now  an example of a point set $S$ with depth $n/3$, in
which the insertion of a suitable point modifies the depth of a
certain point from $f=n/3$ to $2$. Let us consider two triangles
homothetic from their common circumcenter such that the
circumcircle $C$ of the inner triangle $T_{int}$crosses twice each
edge of the outer triangle $T_{ext}$(see Figure \ref{fig6.24}).
Then $S$ is defined by taking the six vertices of the triangles
and placing evenly points in the segments $s_1$ $s_2$ and $s_3$)
that join corresponding vertices of both triangles. Notice that
the interior of the disk bounded by $C$ is empty of points of $S$
and that part of it is outside $CH(S)$. The Delaunay layers of $S$
are triangles and the depth of $S$ is $n/3$; layers and levels are
shown in Figure \ref{fig.localiz} (top).

\begin{figure}[ht]
\begin{center}
\includegraphics[width=0.4\columnwidth]{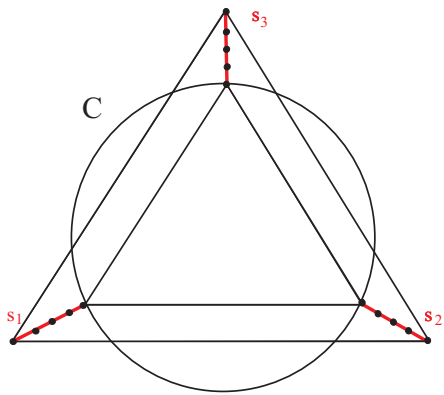}
\caption{The points of $S$ lie on the segments $s1$, $s_2$ and $s_3$.}\label{fig6.24}
\end{center}
\end{figure}

We insert now a point $p$ (refer to Figure \ref{fig.localiz})
which is exterior to $CH(S)$ and interior to the disk bounded by
$C$. In this way, $p$ is adjacent to the three vertices of
$T_{int}$ and to all points placed on the two closest segments
$s_i$, let them be, for example, $s_1$ and $s_2$. Hence $p$ is
adjacent to points of depth $n/3$ in $S$ (the vertices of
$T_{int}$) and to points of depth $1$ (the vertices of $T_{ext}$).

Let us compute the depths in the  $S\cup \{p\}$. The point $p$ has
depth $1$ (it is exterior to $CH(S)$) and any of its neighbors
that is not in that hull has now depth $2$. Therefore, at least
one point of depth equal to $n/3$ in $S$, has depth $2$ in $S\cup
\{p\}$, a change as claimed.

The points of depth $1$ and $2$ in $S$ have still the same depth
in $S\cup\{p\}$. The edges of $DT(S\cup\{p\})$ with an endpoint in
$s_3$ are the same as in $DT(S)$; only edges between $s_1$ and
$s_2$ have changed. As a consequence, the point of $Lay_2^D(S)$
from $s_3$  and the neighbors of $p$ in $Lay_2^D(S\cup\{p\})$
determine a cycle of $Lay_2^D(S\cup\{p\})$ (Figure
\ref{fig.localiz}, bottom, left). The other points that remain on
$s_3$ are of depth $3$. Therefore, after the insertion of $p$, de
depth of $S$ changes from $n/3$ to $3$.

\begin{figure}[ht]
\begin{center}
\includegraphics[width=0.8\columnwidth]{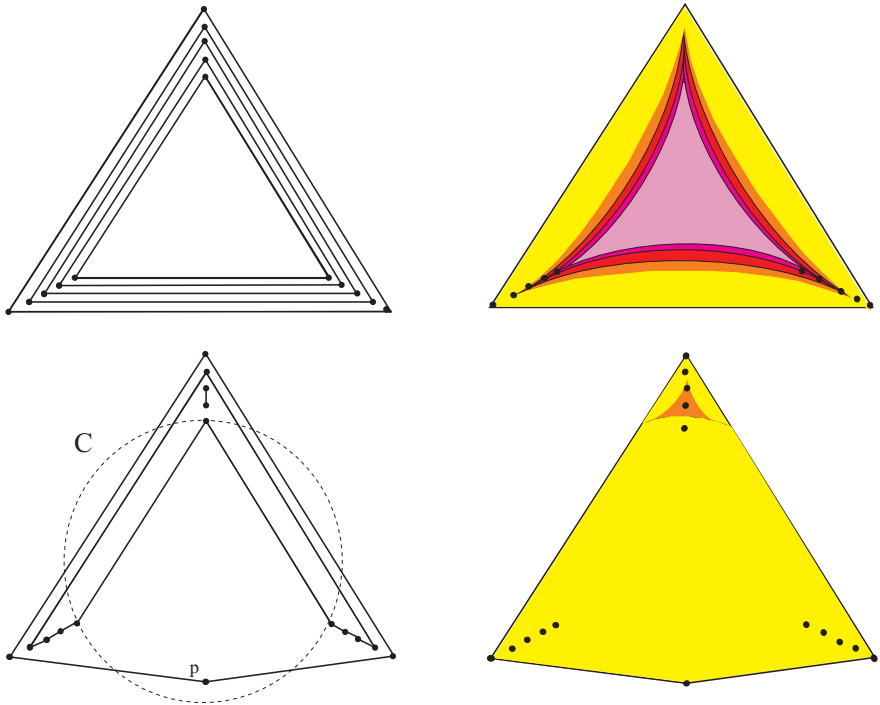}
\caption{Delaunay layers and levels of the sets of points $S$ (at the top) and $S\cup \{p\}$ (at the bottom).}
\label{fig.localiz}
\end{center}
\end{figure}

$\hfill \Box$
\end{proof}

\section{Conclusion}
In this work we have studied the Delaunay depth function, the
stratification that this depth induces in the point set (layers)
and in the whole plane (levels), and developed algorithms for
computing the \emph{Delaunay depth contours} and the depth of any
query  point set with respect to the given point set. The
stratification suggests that Delaunay depth may be more suitable
than others for cluster detection and visualization.

As for open problems, let us mention that we don't know whether a
Delaunay median, i.e., a point of maximal depth, can be computed
directly, escaping depth computation for the whole point set.

\end{document}